\documentclass[12pt]{elsart}
\usepackage{epsfig}
\usepackage{amssymb}

\begin{document}

\begin{frontmatter}



\title{Quark-quark double scattering and modified (anti-)quark fragmentation
functions in nuclei}


\author[CCNU,Regensburg,TAMU]{ Ben-Wei Zhang},
\author[LBL]{ Xin-Nian Wang} and
\author[Regensburg]{Andreas Sch\"afer}

\address[CCNU]{Institute of Particle Physics, Huazhong Normal University,
         Wuhan, China}
\address[Regensburg]{Institut f\"ur Theoretische Physik, Universit\"at
Regensburg, Regensburg, Germany}
\address[TAMU]{Cyclotron Institute, Texas A$\&$M
University, College Station, TX, USA}
\address[LBL]{Nuclear Science Division,
Lawrence Berkeley National Laboratory, CA, USA}

\begin{abstract}
Quark-quark double scattering in eA DIS and its contribution to
quark and anti-quark fragmentation functions are investigated with
the generalized factorization of the relevant twist-four processes
in pQCD. It is shown that the resulting modifications to quark and
anti-quark fragmentation functions are different. While the
numerical size of these effects cannot be determined from pQCD,
the structure of our result leads to a number of qualitative
predictions for the relative size of the effect for different hadrons.
These qualitative predictions agree with the multiplicity ratios
for positive and negative hadrons as observed by HERMES.
\end{abstract}

\end{frontmatter}


Multiple scattering in a nuclear medium is a key process to
understand how hard processes are modified in nuclei. Detailed
experimental information, which is available for such diverse
reactions as electron-nucleus collisions ($e+A$) in fixed target
kinematics and hadron-nucleus collisions ($h+A$) and nucleus-nucleus
collisions ($A+A$) at RHIC or LHC is confronted by highly advanced
theoretical descriptions \cite{gvwz,Qiu:2001,GW,ZW,BSZ,wang03}.
Still, however, a truly comprehensive and reliable description of
all phenomena has not been reached. In this contribution we discuss
one element which should be contained in such an ultimate
theoretical description.

We analyse \cite{ZWS06} quark-quark double scattering in nuclei and
consider the nuclear effects on the quark and anti-quark
fragmentation functions (FF), using the generalized factorization
theory by LQS \cite{LQS} in perturbative QCD. We find that
quark-quark double scattering will give different corrections to
quark and anti-quark FF, and obtain qualitative predictions which
are in agreement with quite remarkable observation by the HERMES
Collaboration \cite{hermes}.

\begin{figure}
\centerline{\psfig{file=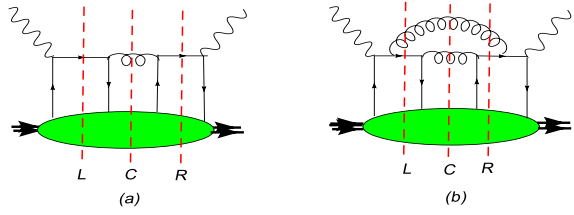,width=3.in,height=1.1in}}
\caption{Two typical diagrams for quark-quark double scattering with
three possible cuts, central(C), left(L) and right(R): (a)
lowest-order of quark-quark rescattering; (b)next-to-leading order
quark-quark rescattering.} \label{fig:4q-EX}
\end{figure}

In general, there are two different kinds of double scattering with
nuclear enhancement in eA DIS: quark-gluon double scattering and
quark-quark double scattering. Quark-gluon double scattering gives
the dominant contribution to the leading quark energy loss and has
been studied in detailed \cite{GW,ZW,EW1}, while quark-quark double
scattering may mixes quark and gluon fragmentation functions and
therefore gives rise to new nuclear effects, as pointed out by Wang
and Guo \cite{GW}. This is illustrated in Fig.~\ref{fig:4q-EX}(a).
For the central-cut diagram, the final parton is a gluon fragmenting
into the hadron considered. For the left-cut or right-cut diagrams,
however, the final fragmenting parton is a quark. Therefore, adding
up the contributions of these three cut diagrams we will get a
mixture of quark and gluon fragmentation functions.  Moreover, we
can see from the central cut of diagram (a) that the quark jet is
converted into a gluon jet by annihilation with an anti-quark in the
nucleus, which provides a mechanism to change the jet flavor
\cite{LKZ}.

For quark-quark double scattering, there are three cut diagrams in
leading-order as shown in Fig.~\ref{fig:4q-EX}(a), and 46 cut
diagrams in next-to-leading order, see Fig.~\ref{fig:4q-EX}(b)
\cite{ZWS06}. By summing all these contributions and including
virtual corrections we can rewrite the semi-inclusive tensor in
terms of a modified fragmentation function taking into account
quark-quark double scattering:
\begin{eqnarray}
\frac{dW_{\mu\nu}}{dz_h}=\sum_q \int dx {f}_q^A(x,\mu_I^2)
H^{(0)}_{\mu\nu} \widetilde{D}_{q \to h}(z_h,\mu^2) \label{eq:Wtot}
\\
\widetilde{D}_{q\to h}(z_h,\mu^2)= D_{q \to h}(z_h,\mu^2)+ \Delta
\widetilde{D}_{q\to h}(z_h,\mu^2)\,\, , \label{eq:mff-1}
\end{eqnarray}
where $H^{(0)}_{\mu\nu}$ is the partonic part for single scattering,
$D_{q\to h}(z_h,\mu^2)$ are the leading-twist quark fragmentation
functions in the vacuum, and the $\Delta \widetilde{D}_{q\to
h}(z_h,\mu^2)$ are related to the twist-four parton correlation
operators \cite{GW,ZW,LQS} and could be approximated as:
\begin{eqnarray}
& &\Delta \widetilde{D}_{q\to h}(z_h,\mu) \approx\frac{2\pi\alpha_s
x_B}{x_A Q^2} \frac{2C_F}{N_C} [D_{g\to h}(\frac{z_h}{z})-D_{q\to
h}(\frac{z_h}{z})] C\, f_{\bar q}^N(x_T)
\nonumber \\
& & + \int_0^{\mu^2} \frac{d\ell_T^2}{\ell_T^2} \frac{\alpha_s^2
x_B}{x_A Q^2} \int_{z_h}^1 \frac{dz}{z} D_{q \to h}(\frac{z_h}{z})
\frac{1+z^2}{(1-z)^2_+} \nonumber \frac{C_F}{N_C}
[1-e^{-x_L^2/x_A^2}] C\, f_{\bar q}^N(x_T)
\nonumber \\
& & -\int_0^{\mu^2} \frac{d\ell_T^2}{\ell_T^2} \frac{\alpha_s^2
x_B}{x_A Q^2} \int_{0}^1 dz D_{q \to h}(z_h) \frac{1+z^2}{(1-z)^2}
\nonumber  \frac{C_F}{N_C} [1-e^{-x_L^2/x_A^2}] C\, f_{\bar
q}^N(x_T)\, . \nonumber
\end{eqnarray}
Here $x_A=1/m_N R_A$, $f_q^A(x)$ is the quark distribution inside a
nucleus, $f_{\bar q}^N(x)$ is the anti-quark distribution inside a
nucleon and $C$ is assumed to be a constant, $ x_T=<k_T^2>/2p^+q^-z$
and $k_T$ is the typical transverse momentum carried by the
antiquark. If we consider the modification to an anti-quark FF all
the calculations are the same with  $ q \leftrightharpoons {\bar q}
$. Therefore, we have $$ \frac{\Delta \widetilde{D}_{ q\to
h}(z_h,\mu^2) }{\Delta \widetilde{D}_{\bar q\to
h}(z_h,\mu^2)}\approx \frac{f_{\bar q}^N(x_T) }{f_{q}^N(x_T)} \,\, <
1 \,\, $$ .

HERMES \cite{hermes} has measured the multiplicity ratio $R_M^{h}$,
which represents the ratio of the number of hadrons of type $h$
produced per DIS event for a nuclear target of mass A to that from a
deuterium target (D):
\begin{equation}
R_M^{h}(z,\nu) = { { \frac{N_h^A(z,\nu)}{N_e^A(\nu)} }  \Biggm/
                   { \frac{N_h^D(z,\nu)}{N_e^D(\nu)} }  }
\label{eq:Rm}
\end{equation}
\noindent with $N_h(z,\nu)$ the number of semi-inclusive hadrons in
a given ($z,\nu$)-bin, and $N_e(\nu)$ the number of inclusive DIS
positrons in the same $\nu$-bin.

In the constituent quark model, we have $ p \,\,= \,\, uud \,\,
,\bar p \,\, = \,\, \bar u \bar u \bar d\,\,\,$ , $ \pi^+, \pi^0,
\pi^- \,\, = \,\, u\bar d \,\,\,  , (u\bar u \, - d\bar d\, )/\sqrt
2 \,\, \, , d\bar u \,\,\, $. Because the gluon FF for a positive
hadron and its antiparticle are the same, and as we are interested
only in the difference between the multiplicities of positive
hadrons and negative hadrons, we neglect the contributions from
gluon FF and simply assume:
\begin{equation}
\Delta R_M^{h}(z,\nu)\equiv 1-R_M^{h}(z,\nu) \approx
\frac{1}{N}\sum_a \Delta \widetilde{ D}_{a\to h}(z,\nu)\,\, ,
\label{Rm-app}
\end{equation}
where $a$ is a constituent quark of hadron $h$, and $N$ is the
number of these constituent quarks.

Therefore we obtain
\begin{eqnarray}
& &\Delta R_M^{\pi}  \approx \frac{1}{2}( \Delta\widetilde{D}_{q\to
h}(z,\nu) + \Delta \widetilde{D}_{\bar q\to h}(z,\nu) ) \propto
[f_{\bar q}^N(x_T)+f_{ q}^N(x_T)] \, \, ,
\label{eq:dRm-pion} \\
& &\Delta R_M^{p} \approx  \Delta\widetilde{D}_{q\to h}(z,\nu)
\propto f_{\bar q}^N(x_T) \,\, , \Delta R_M^{\bar p} \approx
\Delta\widetilde{D}_{\bar
q\to h}(z,\nu) \propto f_{ q}^N(x_T) \,\, , \label{dRm-proton} \\
& &\Delta R_M^{\pi^+}(z,\nu) \simeq \Delta R_M^{\pi^-} \simeq \Delta
R_M^{\pi^0} \,\, ,\label{eq:pion-order} \\
& &\Delta R_M^{\bar p}(z,\nu)
> \Delta R_M^{p}(z,\nu) \,\, . \label{eq:proton-order}
\end{eqnarray}

Similarly we get
\begin{eqnarray}
\Delta R_M^{K^-}(z,\nu)  > \Delta R_M^{K^+}(z,\nu) \,\, , \ \ \
\Delta R_M^{h^-}(z,\nu) > \Delta R_M^{h^+}(z,\nu) \,\, \
.\label{eq:hadron-order}
\end{eqnarray}
The qualitative statements of  Eq.~(\ref{eq:pion-order}),
(\ref{eq:proton-order}), and (\ref{eq:hadron-order}) agree with the
observations made by the HERMES collaboration \cite{hermes}.

\section*{Acknowledgements}
The authors thank Jian-Wei Qiu and Enke Wang for helpful discussion.
This work was supported by Alexander von Humboldt Foundation, by
BMBF, by NSFC under project No. 10405011, by MOE of China with
project No. CFKSTIP-704035, the U.S. Department of Energy under
Contract No. DE-AC02-05CH11231, and by the US NSF under Grant No.
PHY-0457265, the Welch Foundation under Grant No. A-1358.

\end{document}